\documentclass[%
 reprint,
 amsmath,amssymb,
 aps,
 prl,
 floatfix
]{revtex4-2}

\usepackage[english]{babel}
\usepackage{graphicx}
\usepackage{dcolumn}
\usepackage{bm}
\usepackage{mathtools}

\makeatletter
\let\originalselectlanguage\selectlanguage
\renewcommand{\selectlanguage}[1]{%
  \def\temp{#1}\def\tempen{en}%
  \ifx\temp\tempen
    \originalselectlanguage{english}%
  \else
    \originalselectlanguage{#1}%
  \fi
}
\makeatother

\begin{document}

\title{Compositional proofreading through critical self-tuning}

\author{Omer Karin}
\affiliation{%
 Department of Mathematics, Imperial College London, London, SW7 2AZ, UK
}%   

\begin{abstract}
High-dimensional multicomponent systems, including immune and epigenetic repertoires, must selectively retain rare, beneficial components while purging a massive influx of suboptimal variants. We demonstrate that critical tuning of component control parameters through competition naturally implements proofreading in these systems. Competition for shared inputs pins the system to the marginal stability threshold of the most persistent species. This grants dominant species extended lifetimes, concentrating the population into dominant components while forcing less-stable variants into rapid drift-driven turnover. When aggregate drive exceeds a characteristic scale, this pinning fails, producing a non-selective state where component lifetimes scale as a universal power law with aggregate drive. Applying this framework to biological memory, we identify the hallmarks of this effect in plasma cell accumulation dynamics and propose that de-pinning transitions may represent failure points across biological domains, including cancer, immunodeficiencies, and the aberrant activation of harmful genomic elements during ageing.
\end{abstract}

\maketitle

\emph{Introduction} -- A fundamental requirement for biological organization is the ability to amplify and retain high-quality, functional components. At the molecular scale, the fidelity of replication, translation, and signal transduction relies on distinguishing correct substrates from a background of chemically similar, but functionally inferior, competitors. The classical solution to this problem is kinetic proofreading \cite{hopfield1974kinetic,ninio1975kinetic,mckeithan1995kinetic,savir2007conformational, murugan2014discriminatory,ravasio2024minimal}, where energy dissipation is coupled to irreversible steps to actively reject incorrect intermediates.

An analogous challenge arises at the macroscopic scale in populations of long-lived components, such as adaptive immune repertoires and silenced genomic domains. These consist of dynamic entities undergoing continuous turnover, with fate decisions governed by high-dimensional networks that selectively retain rare, beneficial components while purging an influx of suboptimal variants. Primary examples include cell-based adaptive immunity \cite{tonegawa1983somatic,radbruch2006competence,altan2020quantitative,simons2024tuning}, epigenetic memory \cite{beltran2020epimutations,karin2023epigenetic}, neural development \cite{purves1980elimination,goodman1993developmental,bentley2009tipping} and stem cell persistence \cite{van2023cell,simons2025cell}. The exploratory nature of these processes is thought to be crucial for conferring robust and evolvable functionality \cite{gerhart2007theory}.

An important question is how the underlying regulatory networks may achieve high-fidelity compositional proofreading without centralised control. The dynamics of component removal in these systems point to a possible mechanism. Decisions regarding survival and death are frequently governed by discontinuous tipping points \cite{bagci2006bistability,albeck2008modeling,yao2011origin,briffa2024dissecting,simons2025cell}. In this dynamical regime, a component may be stable until a control parameter crosses a critical threshold, triggering a rapid, irreversible transition to elimination. Near such critical tipping points, systems exhibit a universal steep dependence of transition rates on underlying parameters, as well as distinctive dynamical properties including critical slowing-down \cite{stanoev2020organization,karin2023epigenetic,koch2024biological,simons2025cell}. A wide range of biological systems, including immune and epigenetic memory, have components that appear to operate near criticality \cite{mora2011biological,karin2023epigenetic,simons2024tuning,simons2025cell}. 

We have recently shown that self-tuning to the vicinity of a critical point naturally emerges from competition over shared factors that couple the control parameters of different components \cite{karin2023epigenetic,simons2024tuning,simons2025cell,agranov2025self}. For example, immune memory cells compete for common survival factors, while epigenetic memory units (chromatin modifications, small RNAs) compete globally over shared amplification machinery. The interplay between stochastic production and coupled removal constrains component turnover through a queue-like mechanism. This process, which we termed self-tuning by competition, provides a generic route for systems to organise near a critical point and generates characteristic system-level responses to perturbations \cite{karin2023epigenetic}.

Here, we propose that critical self-tuning by competition serves as an efficient compositional proofreading mechanism through a pinning transition. To show this, we extend the previous analyses, which focused on homogeneous populations, to heterogeneous natural repertoires consisting of species with varying sensitivities to stabilising signals. We show that competition for shared resources pins the common environment to the critical point of the most stable species. This pinning \cite{fisher1998collective,reichhardt2017depinning} creates dynamical segregation in which stable species persist in a critically slowed state, while less stable species undergo rapid, deterministic removal and are efficiently depleted from steady-state repertoires. Conversely, increased aggregate production can drive the system into a de-pinned regime where proofreading capacity is lost, a transition that may contribute to pathological conditions.

\emph{Model} -- We consider a dynamical system composed of $R$ interacting species. A Poisson process with rate $\lambda_r$ introduces components of species $r$ into the system. The removal mechanism of each component is governed by a high-dimensional regulatory network, whose effective dynamics are captured by the stochastic low-dimensional form (Appendix S1): 
\begin{equation}
dx_{i,r} = \left( x_{i,r}^2 - \mu_r + h(t) \right) dt + \sigma_r \, dW_{i,r},
\label{eq:langevin}
\end{equation}
where the variable $x_{i,r}$ represents the reaction coordinate of the $i$-th component of species $r$, $dW_{i,r}$ denotes a standard Wiener increment, and $\sigma_r$ is the noise amplitude. The intrinsic stability parameter $\mu_r > 0$ sets the baseline distance to the bifurcation point. The global field $h(t)$ integrates the population of active components $N_r(t)$ according to
\begin{equation}
h(t) = \sum_{r=1}^R \gamma_r N_r(t),
\label{eq:field}
\end{equation}
where the coupling constant $\gamma_r\geq0$ quantifies the contribution of species $r$ to the environmental load.

Components are defined as active while they reside within the viable region of the state space, $x < x_{\mathrm{esc}}$. The mean residence time $\tau(z)$, defined as the mean first-passage time to $x_{\mathrm{esc}}$ for a component with effective barrier $z \equiv \mu_r - h$, interpolates between Kramers escape for $z > 0$ and drift-driven transit for $z < 0$ (Appendix S2). We index species by their intrinsic stability such that $\mu_1 < \mu_2 < \dots < \mu_R$. The macroscopic mean residence time (or lifespan) of a species is given by $T_r=\tau(z_r)$.

Equations~(\ref{eq:langevin}) and~(\ref{eq:field}) define a closed feedback loop. Component accumulation drives the global field $h$ upward. This rise reduces the effective barrier $z_r = \mu_r - h$ for all species, accelerating the passage of components through the viable region, either by lowering the escape barrier or by steepening the drift gradient, creating a self-limiting mechanism.

\emph{Single-species model} -- We first analyse the behaviour of a solitary species with stability $\mu$ and self-interaction $\gamma$. The mean residence time $\tau(z)$ depends on the effective barrier $z = \mu - \gamma N$. The steady-state population $N$ is then determined by the balance between production and removal,
\begin{equation}
N = \lambda \, T.
\label{eq:balance}
\end{equation}

\begin{figure}[t]
\centering
\includegraphics[width=\linewidth]{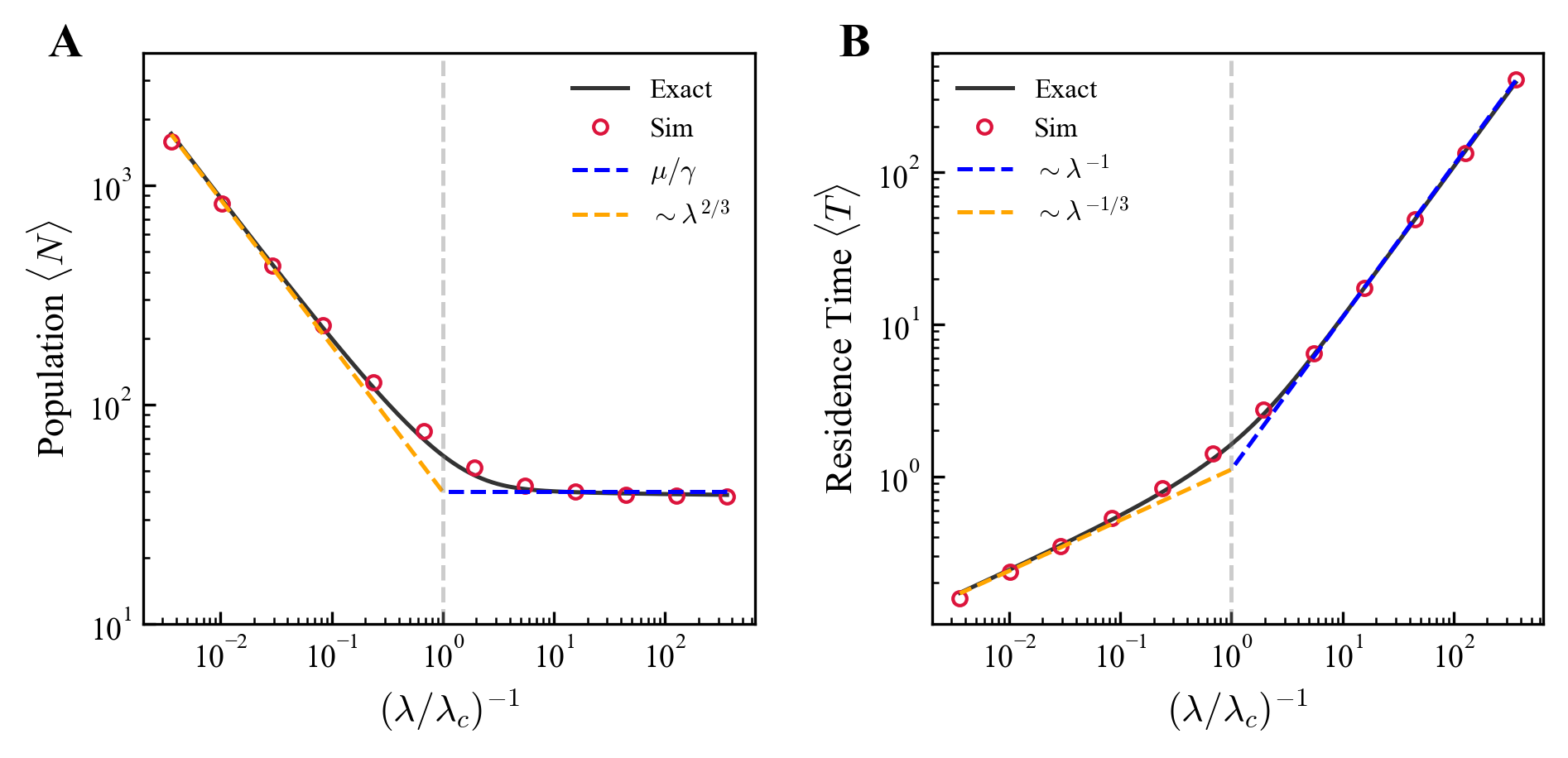}
\caption{\textbf{Transition from pinned to drift-driven regimes in a single species.} Steady-state statistics for a solitary species ($\mu=2.0$, $\gamma=0.05$) as a function of the normalized inverse drive $(\lambda/\lambda_c)^{-1}$. \textbf{(A)}~Mean population $\langle N \rangle$. The system exhibits a sharp crossover at the critical production rate $\lambda_c$ (vertical dashed line). Under sub-critical drive (pinned regime, right), the population pins to the capacity $N \approx \mu/\gamma$. Under super-critical drive (left), the barrier collapses and the population scales as $N \propto \lambda^{2/3}$. \textbf{(B)}~Mean residence time $\langle T \rangle$. Open circles represent stochastic simulations; solid lines indicate exact analytical theory.}
\label{fig:single_species}
\end{figure}

This governing equation reveals a crossover between two distinct dynamical regimes, separated by a characteristic production rate $\lambda_c = 2 \mu^{3/2} / (\pi \gamma)$ (Appendix S3). In the \textit{pinned regime} ($\lambda < \lambda_c$), the population increases until the effective barrier vanishes ($z \to 0^+$). The occupancy is pinned to the capacity $N \approx \mu \gamma^{-1}$ (Figure~\ref{fig:single_species}A), independent of the production rate $\lambda$. Consequently, the residence time scales inversely with drive, $T \approx \mu \gamma^{-1} \lambda^{-1}$ (Figure~\ref{fig:single_species}A).

Conversely, in the \textit{drift-driven} (super-critical) regime ($\lambda > \lambda_c$), the drive is sufficient to collapse the barrier entirely ($z < 0$). The population is then limited only by the transit speed through the saddle-node region. Using the scaling $\tau(z) \approx \pi/(2\sqrt{|z|})$, we derive the population scaling $N \approx \mu \gamma^{-1}(\lambda/\lambda_c)^{2/3}$ (see Supplemental Material). Crucially, the residence time exhibits a crossover to a power-law decay, 
\begin{equation}
T \approx T_c \left( \frac{\lambda}{\lambda_c} \right)^{-1/3},
\end{equation}
where $T_c = \pi/(2\mu^{1/2})$ is the critical residence time at the onset of the super-critical regime.

\begin{figure}[t]
\centering
\includegraphics[width=\linewidth]{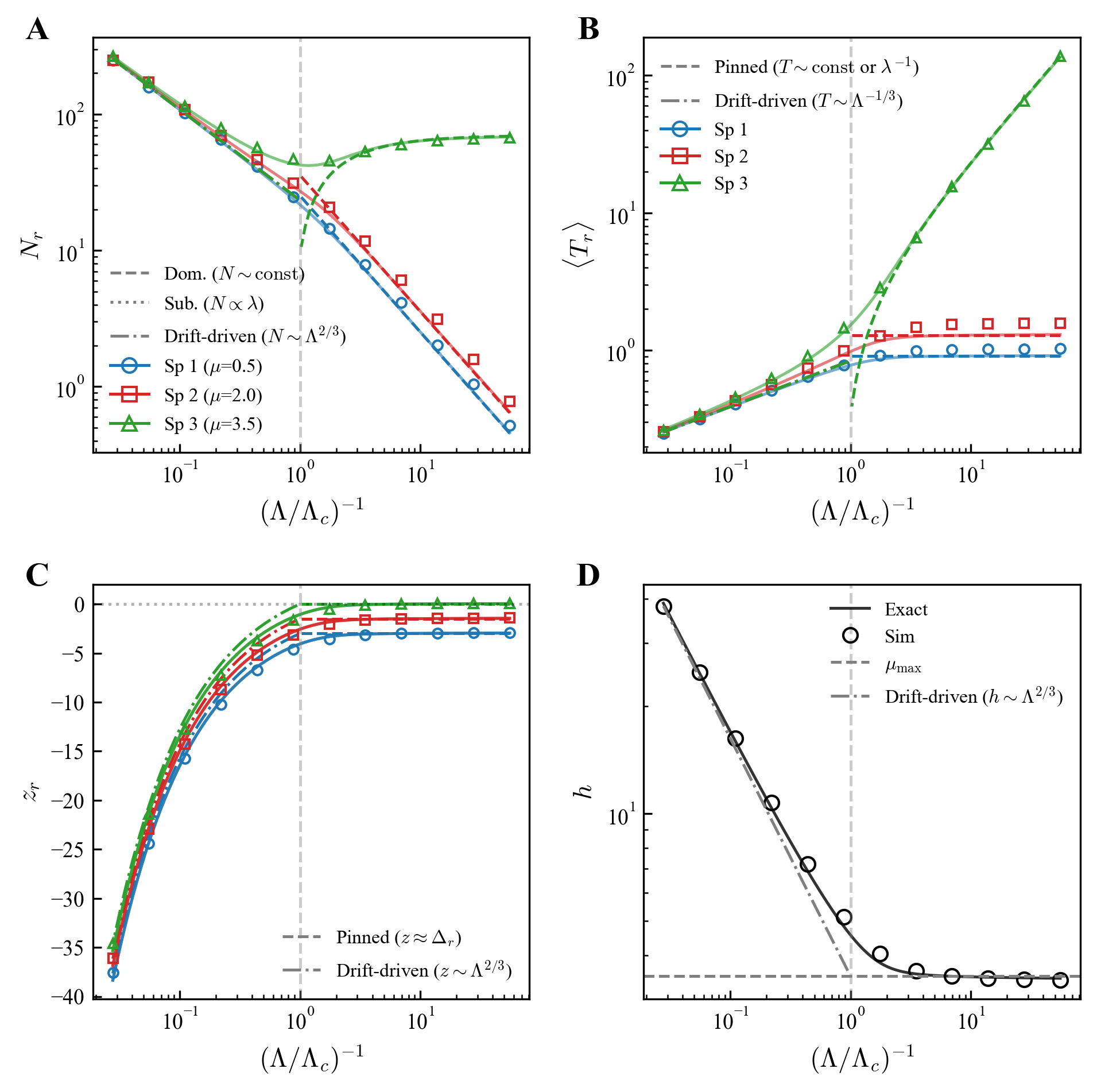}
\caption{\textbf{Compositional proofreading through competitive self-tuning.} Steady-state behavior of $R=3$ interacting species with distinct stabilities ($\mu \in \{0.5, 2.0, 3.5\}$) competing via a shared global field, plotted against the normalized inverse aggregate drive $(\Lambda/\Lambda_c)^{-1}$. \textbf{(A)}~Species populations $N_r$. Sub-dominant species (blue, orange) follow linear scaling, while the dominant species (green) buffers the system capacity. \textbf{(B)}~Residence times $T_r$. The dominant species achieves extended longevity near the pinning transition ($\Lambda \approx \Lambda_c$, vertical dashed line), whereas sub-dominant lifetimes remain short and constant. In the drift-driven regime (left), all component lifetimes collapse universally. \textbf{(C)}~Effective barriers $z_r$. The field pins such that the dominant barrier approaches zero, leaving sub-dominant species with finite instability gaps. \textbf{(D)}~Global field $h$. The field is clamped to $h \approx \mu_{\max}$ in the pinned regime but grows as $\Lambda^{2/3}$ in the drift-driven regime. Symbols denote stochastic simulations; solid lines denote analytical theory.}
\label{fig:multi_scaling}
\end{figure}

\emph{Proofreading in multi-species competition} --
We now generalize to $R$ interacting species coupled via the global field $h$. The steady-state is determined by the self-consistency conditions on individual species, $N_r=\lambda_r \, \tau(z_r)$, from which we derive the self-consistency condition on the input field:
\begin{equation}
h = \sum_{r=1}^R \gamma_r \lambda_r \, \tau(\mu_r - h).
\label{eq:consistency}
\end{equation}
The system's macroscopic state is controlled by the aggregate interaction flux, $\Lambda \equiv \sum_{r} \gamma_r \lambda_r$. The competition between this aggregate drive and the maximum barrier in the repertoire, $\mu_{\max} \equiv \max_r(\mu_r)$, induces a transition at the critical flux (Figure~\ref{fig:multi_scaling}, Appendix S3):
\begin{equation}
\Lambda_c = 2\mu_{\max}^{3/2}/\pi ,
\label{eq:critflux}
\end{equation}
 between a pinned regime ($\Lambda < \Lambda_c$) and a drift-driven regime ($\Lambda > \Lambda_c$).

For sub-critical drive ($\Lambda < \Lambda_c$), the self-consistency condition (Eq.~\eqref{eq:consistency}) demands that the global field be pinned near the largest barrier: $h \approx \mu_{\max}$.
This pinning creates a sharp timescale separation between the dominant species ($r^*$, where $\mu_{r^*} = \mu_{\max}$) and all sub-dominant species ($r \neq r^*$).
The sub-dominant species perceive a finite effective barrier $\Delta_r = \mu_{\max} - \mu_r > 0$. Their populations are linear in their production rates:
\begin{equation}
N_r \approx \lambda_r\,\tau(-\Delta_r) \quad (r \neq r^*).
\end{equation}
In contrast, the dominant species fills the remaining capacity. Its population is determined not by its production rate, but by the space left by the sub-dominant species:
\begin{equation}
N_{r^*} \approx \frac{1}{\gamma_{r^*}} \left( \mu_{\max} - \sum_{r \neq r^*} \gamma_r N_r \right).
\end{equation}
This mechanism grants the dominant species an effective lifetime $T_{r^*} \propto \lambda_{r^*}^{-1}$ that can be orders of magnitude larger than the intrinsic timescales of the sub-dominant components. Unlike a non-competitive system where representation is determined solely by influx $N_r\propto \lambda_r$, critical pinning non-linearly amplifies the presence of the most stable species while suppressing even high-flux suboptimal variants. This condensation \cite{knebel2015evolutionary} of the population into the dominant species whose abundance scales $N_r\sim \mathcal{O}(N)$, triggered by the pinning of the global field $h$, is the fundamental mechanism driving compositional proofreading.

When the aggregate flux exceeds the critical threshold, the pinning mechanism fails. The global field grows as $h \approx \mu_{\max} (\Lambda/\Lambda_c)^{2/3}$, collapsing the potential barriers for all species.
In this regime, the distinction between dominant and sub-dominant species vanishes. The system flows in a drift-driven manner, and the lifetimes of all components degrade universally according to the power law:
\begin{equation}
T \approx \frac{\pi}{2\sqrt{h}} \sim \Lambda^{-1/3}.
\end{equation}

These results identify self-tuning by competition as a proofreading mechanism acting on steady-state composition. Competition pins the shared environment to the marginal stability of the most persistent species, concentrating occupancy in maximally stable components while depleting less stable ones. Heterogeneity in intrinsic stability is thereby converted into a sharp separation in representation. When the aggregate drive exceeds the pinning capacity, this selectivity is lost and the repertoire becomes non-discriminating as all components enter a drift-driven turnover regime.

\emph{Sequential assembly} -- The compositional proofreading mechanism described so far concerns the properties of the system at steady state. We therefore proceed to examine the system's relaxation to steady state from an initially empty configuration, $N_r(0)=0$ for all $r$. 

\begin{figure}[t]
\centering
\includegraphics[width=\linewidth]{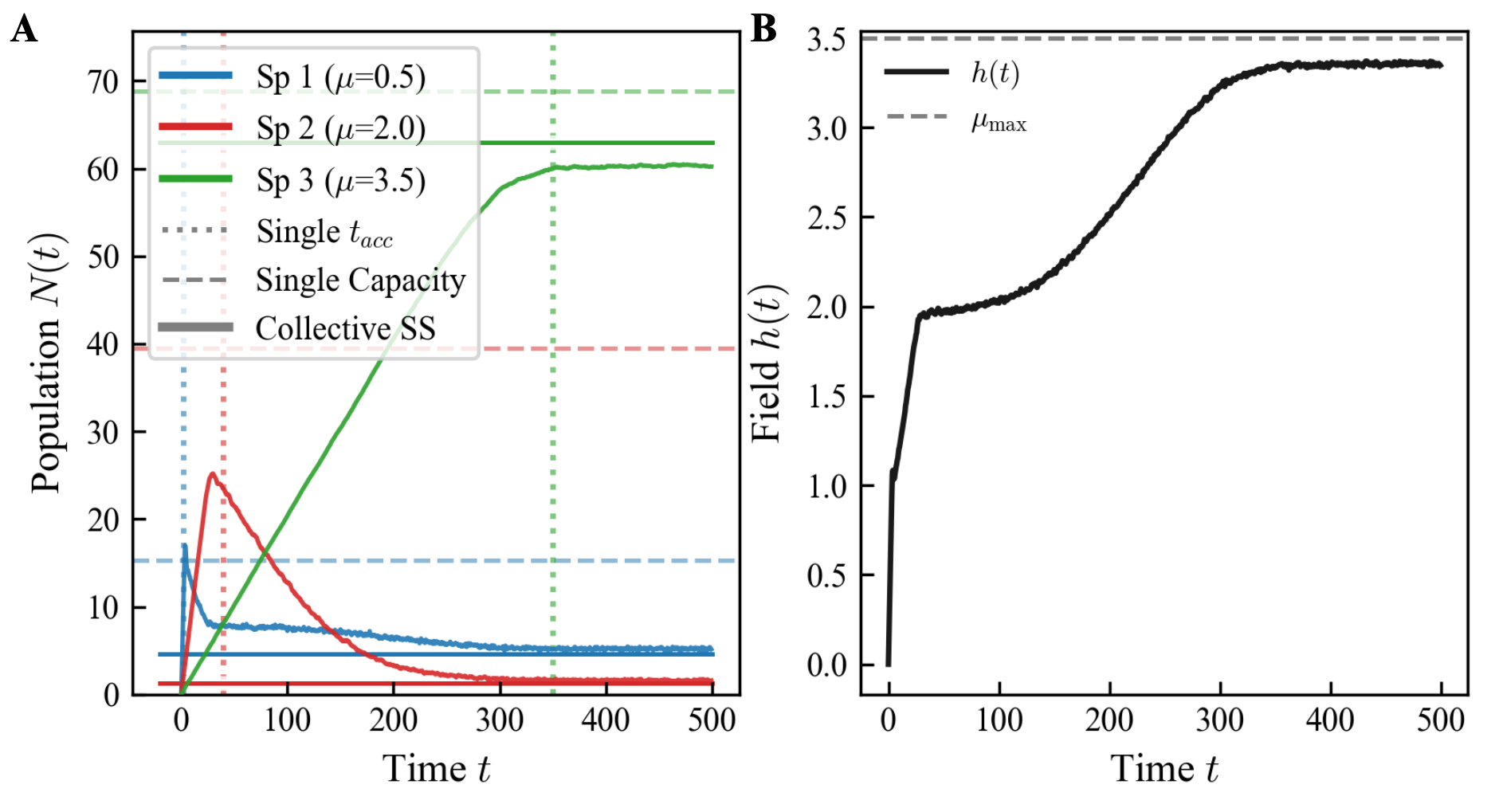}
\caption{\textbf{Sequential assembly and kinetic instability.} Temporal evolution of the system during assembly from an initially empty state under a fixed hierarchy of influx rates ($\lambda \in \{5.0, 1.0, 0.2\}$). \textbf{(A)}~Species populations $N_r(t)$. Fast-accumulating, low-stability components (blue) rapidly overshoot their ultimate drift-driven capacity, triggering a ``sawtooth'' collapse as the environment shifts. This instability progressively filters the repertoire until the most stable species (green) establishes dominance. Dotted vertical lines indicate single-species accumulation timescales $t_{\mathrm{acc}} = \mu_r/(\gamma \lambda_r)$. \textbf{(B)}~The global field $h(t)$. Despite the sawtooth-like collapse of the sub-dominant populations, the macroscopic field rises monotonically.}
\label{fig:accumulation}
\end{figure}

For a single species, the effective barrier is initially large ($z = \mu > 0$), suppressing decay. The population grows linearly, $N(t) \approx \lambda t$, until it reaches the critical threshold at the accumulation timescale $t_{\mathrm{acc}} \approx \mu /(\gamma\lambda)$. In the multi-species case, all populations initially grow linearly, driving a collective, monotonic increase in the global field $h(t) = \sum_{r} \gamma_{r} N_{r}(t)$ (Appendix S4). However, this collective accumulation is punctuated by a sequence of instabilities that result in a sawtooth-like trajectory for the individual populations (Figure~\ref{fig:accumulation}A, Appendix S5).

To gain analytical insight into this sawtooth-like trajectory, we consider the limiting case of strong separation of accumulation timescales between species. In this limit, a subcritical species $k$ initially approaches its single-species capacity, $N_{k}^{\mathrm{peak}} \approx \mu_{k}/\gamma$, before the collective field triggers the bifurcation. Upon crossing the threshold, the population relaxes to a drift-driven steady state, $N_{k}^{\mathrm{ss}} \approx \lambda_{k} \, \tau(-\Delta_k)$. 

While the individual population $N_k$ collapses, the global field $h(t)$ remains monotonic (Figure~\ref{fig:accumulation}B). The collective field is sustained by the continued linear accumulation of the more stable species ($j > k$). However, the loss of species $k$ reduces the aggregate driving flux contributing to the growth of $h(t)$ to $\Lambda_{k+1} = \sum_{j=k+1}^{R} \gamma_{j} \lambda_{j}$. The time required to drive the field from $\mu_{k}$ to the next barrier $\mu_{k+1}$ is thus
$
    t_{\mathrm{transit},k} \approx \frac{\mu_{k+1} - \mu_{k}}{\Lambda_{k+1}}.
$
The global field therefore progressively purges low-stability components via rapid discharge, leaving only the most stable species to drive the system toward its final steady state.

\emph{Repertoire remodelling in plasma cells} -- We apply the proofreading framework to analyse the maintenance of humoral immunity. Long-lived plasma cells (LLPCs), which reside in the bone marrow and may persist and secrete antibodies for years to decades in humans, are generated alongside a vastly larger population of short-lived plasma cells (SLPCs) following infection or vaccination \cite{radbruch2006competence,simons2024tuning}. Although the flux of SLPCs into the marrow exceeds that of LLPCs by orders of magnitude ($\lambda_s \gg \lambda_l$), the steady-state marrow repertoire is dominated by long-lived cells ($N_l > N_s$) \cite{brynjolfsson2018long,jing2024fine}.

We model the bone marrow plasma cell niche as a two-species competitive system, as all plasma cells are coupled through competition over shared survival factors \cite{radbruch2006competence,benet2021plasma,simons2024tuning,eslami2024unique,park2025cd138}. Based on physiological estimates of immune event frequencies and cellular turnover \cite{simons2024tuning}, we parametrize the production fluxes as $\lambda^{\text{hum}}_l \sim 10^6$ and $\lambda^{\text{hum}}_s \sim 10^7$ cells per natural time unit, where the timescale of drift-driven turnover maps to a physiological duration of days to weeks. We set $\mu_l = 1$ and $\mu_s = 1 - \Delta$, where $\Delta = \mu_l - \mu_s > 0$ is the stability gap. As the healthy human niche has a typical size of $N^{\text{hum}}_{\mathrm{niche}} \approx 10^9$ cells, which is orders of magnitude larger than the physiological influx, the system is expected to operate in the pinned regime. We therefore fix $\gamma^{\text{hum}} = \mu_l / N^{\text{hum}}_{\mathrm{niche}} = 10^{-9}$, which yields an aggregate interaction flux of $\Lambda^{\text{hum}} = \gamma^{\text{hum}}(\lambda^{\text{hum}}_s + \lambda^{\text{hum}}_l) \approx 10^{-2}$. The corresponding parameters in mice \cite{simons2024tuning} are $\lambda^{\text{mice}}_l \sim 10^4$ and $\lambda^{\text{mice}}_s \sim 10^5$ with $N^{\text{mice}}_{\mathrm{niche}} \approx 10^6$ cells, setting $\gamma^{\text{mice}}=10^{-6},\, \Lambda^{\text{mice}} \approx 10^{-1}$. Comparing these to the critical flux $\Lambda_c = 2\mu_l^{3/2}/\pi$, we confirm that the healthy bone marrow safely resides in the pinned proofreading state for both mice and humans.

\begin{figure}[t]
\centering
\includegraphics[width=\linewidth]{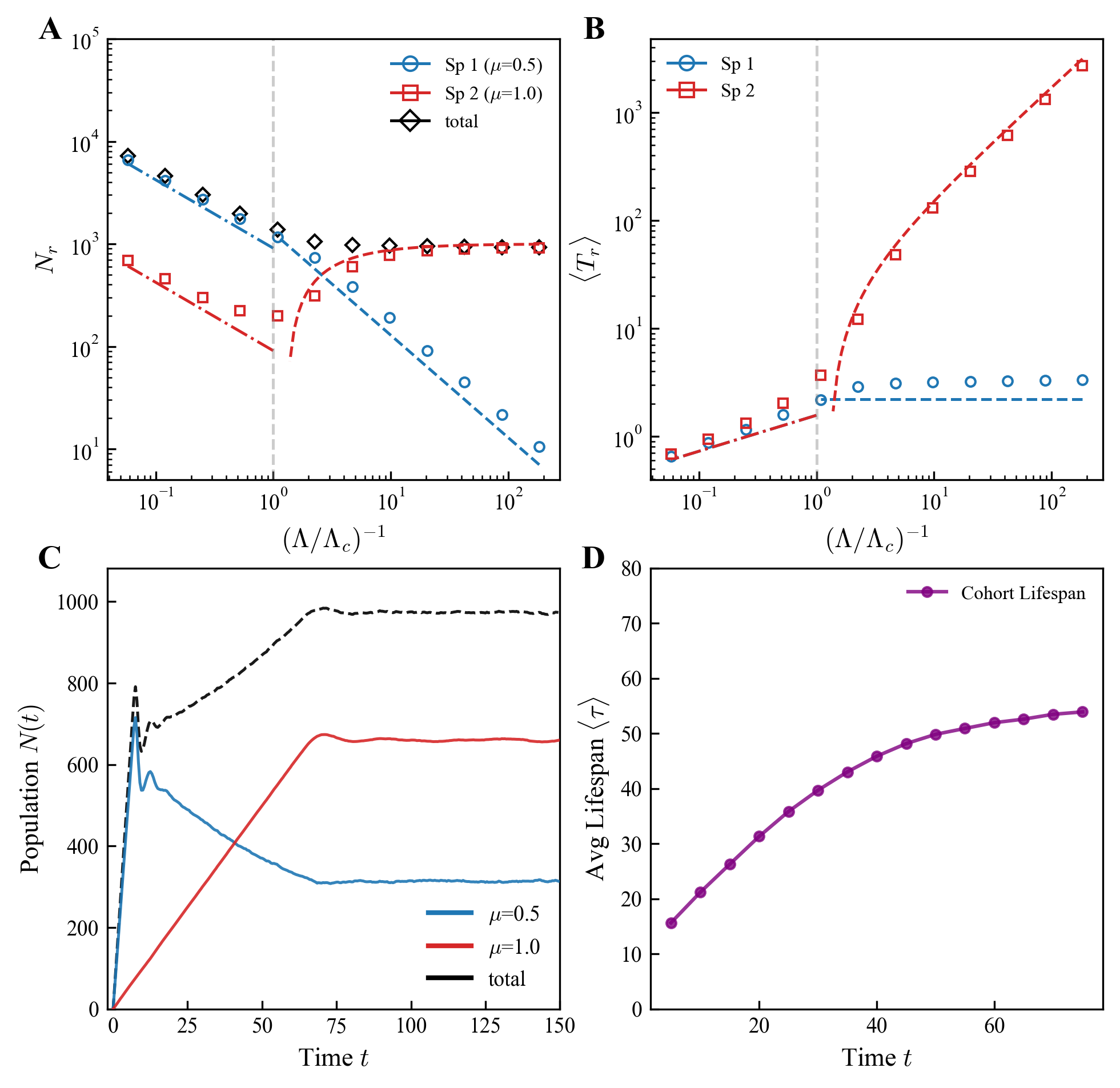}
\caption{\textbf{Transitions and assembly dynamics in a murine plasma cell niche.} Simulations of a two-species system of short-lived ($\mu_s=0.5$) and long-lived ($\mu_l=1.0$) plasma cells. Parameters map to a representative micro-niche encompassing $10^{-3}$ of the physiological mouse bone marrow ($N_{\mathrm{niche}} = 10^3$, $\gamma = 10^{-3}$), conserving the macroscopic flux $\Lambda$. \textbf{(A)}~Steady-state populations and \textbf{(B)}~mean residence times versus inverse aggregate drive $(\Lambda/\Lambda_c)^{-1}$. Sub-critical flux results in environmental pinning, actively selecting for the long-lived state, while super-critical flux triggers a de-pinning transition into the drift-driven failure regime. \textbf{(C)}~Assembly dynamics from an empty niche reveal a two-phase remodeling, involving a rapid influx of short-lived cells, followed by their collapse and the subsequent linear accumulation of long-lived cells. \textbf{(D)}~The average lifespan of plasma cell cohorts progressively extends as the niche remodels into the memory phenotype, reproducing the dynamics observed in murine genetic timestamping experiments.}
\label{fig:plasma_cells}
\end{figure}

In the pinned regime, the steady-state populations decouple from their production rates. The dominant LLPC population fills the niche to its capacity, $N_l/N  \approx 1$, while the sub-dominant SLPC population is tightly suppressed (Figure~\ref{fig:plasma_cells}A). The residence times reflect this separation, with $T_s \sim \mathcal{O}(1)$~days versus $T_l \approx \mu_l/(\gamma\lambda_l)$, which corresponds to years in humans and months in mice (Figure~\ref{fig:plasma_cells}B).

The model further captures the transient dynamics of repertoire assembly during ontogeny or recovery from lymphopenia (e.g., following measles infection or chemotherapy \cite{de2012measles,mina2015long}) (Figure~\ref{fig:plasma_cells}C). Starting from an empty niche, the global field grows rapidly as $h(t) \approx \gamma\lambda_s \, t$. This early phase is dominated by the high-flux SLPCs, which subsequently collapse due to the sawtooth instability described previously. The purging of these lower-stability components facilitates the slow, linear accumulation of LLPCs over a timescale $T_l$ (months in mice, years in humans), progressively extending the average plasma cell lifespan (Figure~\ref{fig:plasma_cells}D). These dynamics quantitatively align with characterisation of the bone marrow niche in mice of different ages and following perturbations, including using genetic timestamping \cite{xu2020genetic,benet2021plasma,liu2022heterogeneous,koike2022progressive,robinson2023intrinsically,jing2024fine}. These studies reveal that there is an initial increase in SLPC numbers, followed by a progressive, months-long transition towards the LLPC-enriched niche phenotype, which occurs in tight association with mouse age and plasma cell number as well as with a progressive extension of plasma cell lifespan, and which is recapitulated after plasma cell depletion. The model thus identifies this two-phase remodelling as a key dynamical signature of proofreading through competition.

The biological relevance of this proofreading mechanism is highlighted by its distinct failure modes in immune pathologies. WHIM syndrome, for instance, presents a paradoxical immunodeficiency characterized by large-scale production of plasma cells and high marrow cellularity, yet poor long-term memory \cite{biajoux2016efficient,gulino2004altered,mc2010oligoclonality,simons2024tuning}. Our framework naturally captures this as a forced de-pinning transition driven by super-critical flux ($\Lambda > \Lambda_c$). As the effective barriers collapse, the system enters the non-selective drift-driven regime where memory lifetime universally degrades as $T \propto \Lambda^{-1/3}$, reproducing the crowded but rapidly turning-over WHIM phenotype. Alternatively, multiple myeloma exemplifies failure via a shifted pinning boundary. Multiple myeloma is characterized by the clonal expansion of neoplastic plasma cells that are hypersensitive to survival signals, and it typically results in severe immunoparesis (suppression of healthy antibodies) \cite{tai2016april,chahin2022clinical}. Within our framework, this malignancy reflects the emergence of a hyper-dominant clone ($\mu_{\mathrm{cancer}} > \mu_l$). By pinning the global field $h$ above the stability threshold of healthy LLPCs, the malignant clone forces the previously stable repertoire into the drift-driven regime, actively purging pre-existing immunity.

\emph{Conclusion} -- We have proposed a compositional proofreading mechanism fundamentally distinct from classical kinetic proofreading. Whereas classical models rely on biochemical kinetic rates, our framework operates through competition among high-dimensional dynamical memory units coupled via shared control parameters. We have shown that in such systems, critical self-tuning implements proofreading by pinning the shared environment to the marginal stability of the most persistent species, selectively retaining stable components while driving suboptimal variants into a drift-driven regime of rapid turnover. 

A complementary mechanism, operating through differences in barrier-fluctuation variance rather than mean barrier height, can further refine repertoires by linking component activity to global functional outcomes, even when intrinsic stabilities are identical \cite{karin2026learning}.

Our framework reveals an intrinsic speed--fidelity tradeoff in compositional proofreading. While repertoire assembly accelerates with aggregate influx ($t_{\mathrm{acc}} \sim \mu_{\max}/\Lambda$), proofreading fidelity requires $\Lambda < \Lambda_c$, and selectivity degrades continuously as $\Lambda \to \Lambda_c$. This tension mirrors the speed--accuracy tradeoff in classical kinetic proofreading~\cite{hopfield1974kinetic,ninio1975kinetic,banerjee2017elucidating}, where additional irreversible steps sharpen substrate discrimination at the cost of time and energy dissipation. In this setting, the cost of a high-fidelity repertoire is an extended formation timescale.

When the aggregate drive exceeds the critical threshold given by Eq.~\eqref{eq:critflux}, this pinning fails and the system transitions to a non-selective drift-driven regime, resulting in loss of compositional fidelity. As the critical flux is a coarse-grained parameter, this failure can proceed through distinct pathways: an increased influx of memory units (through the $\lambda_r$ parameters), modulation of the coupling to the shared environment (through the $\gamma_r$ parameters), or a reduction in the marginal stability of the most persistent species ($\mu_{\text{max}}$). This failure mode may be relevant across a range of physiological conditions. In developing tissues and in the immune system, competition for survival factors plays a central role in quality control at multiple levels, and our framework may help dissect how perturbations such as mutations or chronic infection lead to dysregulation.

Our mechanism may also bear on ageing. In the genome, repetitive elements and retrotransposons that threaten genomic integrity \cite{wood2016chromatin,de2019line} are silenced through compaction into heterochromatin, maintained by specific biochemical modifications to DNA and chromatin \cite{lippman2004role,dodd2007theoretical,oberdoerffer2008sirt1}. This silencing is inherently stochastic \cite{beltran2020epimutations}, with retrotransposons subject to particularly robust, multi-layered 
silencing mechanisms relative to other genomic regions \cite{wood2016chromatin,seczynska2022genome,lehner2025silencing}. Within our framework, silenced genomic regions correspond to memory units whose stability is set by an input field determined by shared maintenance enzymes such as SIR proteins \cite{oberdoerffer2008sirt1,allshire2018ten}. Elements with more robust silencing, typically distinguished by structural features such as the absence of introns \cite{aljohani2020engineering,seczynska2022genome,lehner2025silencing}, correspond to units with larger marginal stability. Heterochromatin is known to decline with age, possibly reflecting reduced availability of SIR proteins \cite{oberdoerffer2008sirt1,van2014sirt6,wood2016chromatin,korotkov2021sirtuin}, and this decline is associated with aberrant activation of harmful genomic regions \cite{wood2016chromatin,korotkov2021sirtuin,gorbunova2021role}. Within our framework, such a decline corresponds to a reduction in the marginal stability of silenced elements, causing the proofreading mechanism to fail and the system to lose the ability to discriminate harmful from benign elements.

\newpage
\bibliography{bib} 

\end{document}